\documentclass[fleqn,10pt]{wlscirep}
\usepackage[utf8]{inputenc}
\usepackage[T1]{fontenc}
\usepackage{graphicx}
\usepackage{amsmath}
\usepackage{bm}
\usepackage{color}
\usepackage{ulem}
\usepackage{cancel}

\title{Topological magnon modes on honeycomb lattice with coupling textures}

\author[1,2]{Hong Huang}
\author[1]{Toshikaze Kariyado}
\author[1,2,*]{Xiao Hu}
\affil[1]{International Center for Materials Nanoarchitectonics (WPI-MANA), National Institute for Materials Science, Tsukuba 305-0044, Japan}
\affil[2]{Graduate School of Science and Technology, University of Tsukuba, Tsukuba 305-8571, Japan}

\affil[*]{HU.Xiao@nims.go.jp}

\begin{abstract}
Topological magnon modes are expected to be useful for novel applications such as robust information propagation, since they are immune to backscattering and robust against disorder. Although there are several of theoretical proposals for topological magnon modes and growing experimental efforts for realizing them by now, it is still desirable to add complementary insights on this important phenomenon. Here, we propose a new scheme to achieve topological magnon where only nearest-neighbour exchange couplings on honeycomb lattice are necessary. In both ferromagnets and antiferromagnets, tuning exchange couplings between and inside hexagonal unit cells induces a topological state accompanied by a band inversion between $p$-orbital and $d$-orbital like magnon modes. Topological magnon modes appear at the interface between a topological domain and a trivial domain with magnon currents, which counterpropagate depending on pseudospins originated from orbital angular momenta of magnon modes. This mimics the spin-momentum locking phenomenon in the quantum spin Hall effect.
\end{abstract}
\begin{document}

\flushbottom
\maketitle
% * <john.hammersley@gmail.com> 2015-02-09T12:07:31.197Z:
%
%  Click the title above to edit the author information and abstract
%
\thispagestyle{empty}

\section*{Introduction}

Recently topology has became a unified key concept in material sciences \cite{Haldane1988,Kane2005,Bernevig2006,Hasan2010,QiXL2011,WengHM2015}. The most prominent feature of topological systems is the surface or edge states. Because of topological protection, these surface or edge states are immune to back-scattering and robust against disorder, which can be exploited for achieving innovative functionalities. The current intense study of topological systems was ignited by the discovery of topological insulators in quantum electronic solids, but this idea has been developed into bosonic systems and various wave phenomena \cite{Haldane2008,Lu2014,Yang2015,Mousavi2015,Huber2016,Khanikaev2017,Ozawa2019}.

Magnons are quanta of spin-wave excitations in magnetic systems. As quasiparticles, magnons are charge neutral bosons and free of dissipation due to Ohmic heating, thus useful for various applications \cite{Kruglyak2010,Lenk2011,Chumak2015}. Novel features of topological magnon modes protected by bulk topology \cite{Shindou2013,ZhangLF2013,Mook2014,Kim2016,Owerre2016} have also been considered. So far, several possible mechanisms for realizing topological magnon modes are proposed \cite{Shindou2013,Shindou2013_2,WangXS2017,McClarty2018,Joshi2018,Kim2019}. Although these proposals are enlightening, experimental realization and firm confirmation are still not easy. Amongst the known proposals, those regarded as promising are using the Dzyaloshinskii-Moriya (DM) interaction in Kagome lattice \cite{ZhangLF2013,Mook2014} or honeycomb lattice \cite{Kim2016,Owerre2016}. In both cases, the DM interaction is crucial for achieving the nontrivial topology, playing the role analogous to the spin-orbit coupling for electrons in the quantum spin Hall effect \cite{Kane2005,Bernevig2006}. However, having a sufficiently strong DM interaction is not necessarily easy, limiting the experimental realization of topological magnon modes. To the best of our knowledge, though there are some experimental progresses in materials with the DM interaction \cite{Onose2010,Chisnell2015}, firm evidences for topological magnon modes are still lacking.

In this work, we propose a new scheme to achieve topological magnon modes on honeycomb lattice, where nontrivial topology is achieved upon tuning nearest-neighbor (n.n.) exchange couplings, whereas no DM interaction is required. It is known \cite{Neto2009,Fransson2016,Boyko2018} that a ferromagnet on  honeycomb lattice with n.n. exchange couplings exhibits Dirac-type linear magnon dispersions. Generally, a Dirac-type linear dispersion can be a nice starting point to have topologically nontrivial states. Here, we demonstrate that introducing a $C_{6v}$-symmetric texture \cite{WLH2015} in the strength of exchange coupling opens a frequency band gap in the magnon frequency band structure, which yields a band inversion and the nontrivial topology. We also extend this idea to antiferromagnets. Unlike ferromagnets, the dispersion of magnon for an antiferromagnet on honeycomb lattice is doubly degenerate owing to the combination of time-reversal symmetry and inversion symmetry, where no Dirac-type dispersion exists \cite{Boyko2018,Owerre2017_3,Maksimov2016}. We notice that this degeneracy is lifted in the canted antiferromagnetic state caused by an external magnetic field except at the $K$ and $K'$ points, which yields Dirac-type dispersions. Therefore, a topological state can be induced by a coupling texture in the canted antiferromagnet with the same mechanism in ferromagnets. In both ferromagnets and antiferromagnets, topological magnon modes appear at the interface between topological and trivial domains, with directions of magnon currents governed by pseudospins \cite{WLH2015,WuLH2016,Kariyado2017}, which mimics the spin-momentum locking phenomenon in the quantum spin Hall effect \cite{Kane2005,Bernevig2006}.

\section*{Results}

\subsection*{Topological magnon modes in ferromagnets}

\begin{figure}[t]
\centering
  \includegraphics[clip=true,width=0.4\textwidth]{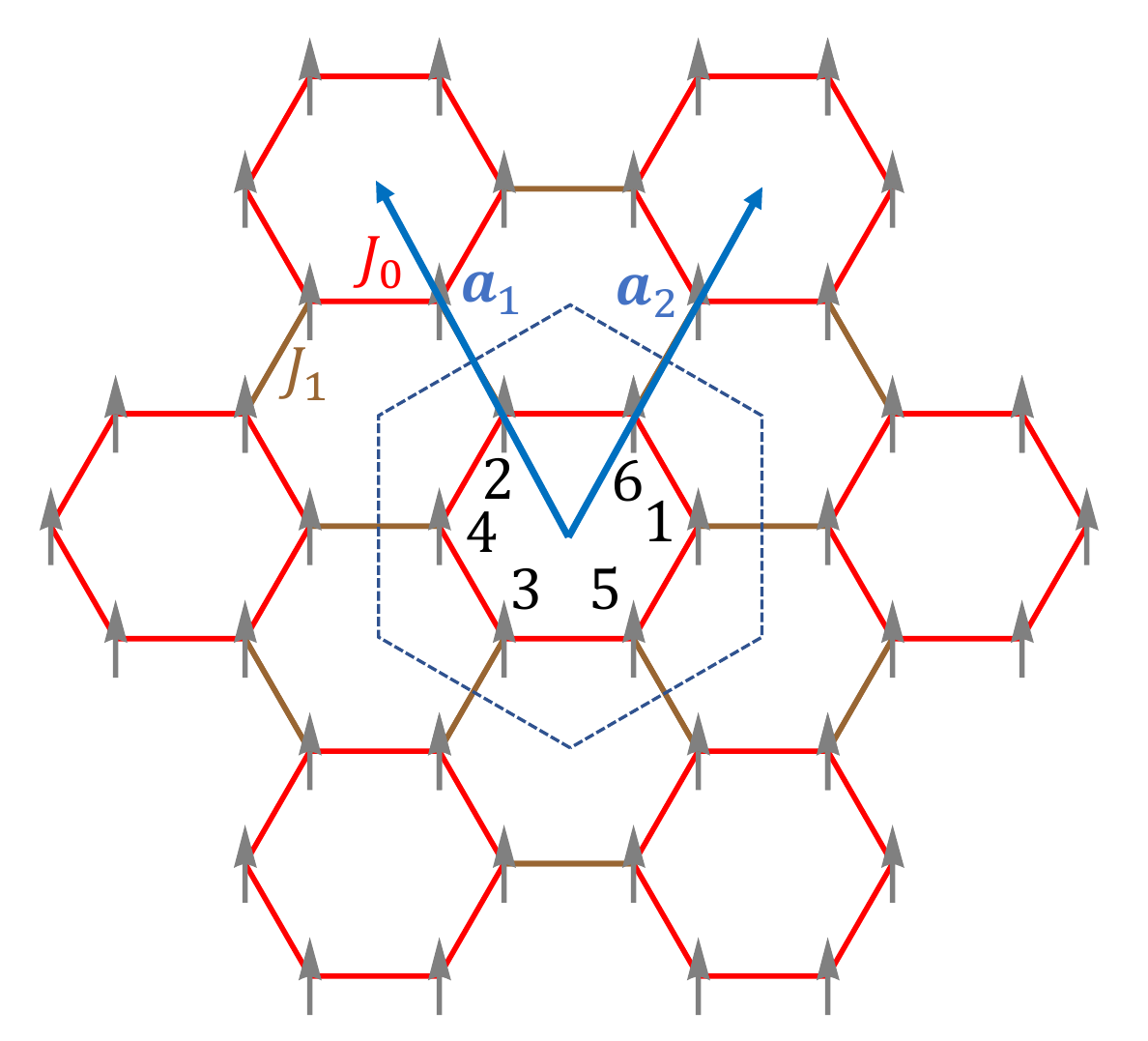}
  \caption{Ferromagnet on honeycomb lattice with nearest-neighbor exchange couplings. Hexagonal unit cells indicated by the dashed line are chosen with $J_0$/$J_1$ representing exchange couplings inside/between hexagonal unit cells, which preserves $C_{6v}$ symmetry. Numbers 1,...,6 index atomic sites inside the hexagonal unit cell. Unit vectors are represented by $\bm{a}_1$ and $\bm{a}_2$, and $|\bm{a}_1|=|\bm{a}_2|=a_0$.}
  \label{Fig:1}
\end{figure}

We start from the Heisenberg model
\begin{equation}
  H=-\sum_{\langle i,j\rangle}J_{ij}\bm{S}_i\cdot \bm{S}_j,
  \label{eqHM} % Heisenberg Model
\end{equation}
where the summation runs over n.n. sites of honeycomb lattice, and $J_{ij}>0$ denote ferromagnetic exchange couplings. As illustrated in Fig.~\ref{Fig:1}, we use a unit cell containing six sites for the convenience of modulation in strength of exchange coupling, instead of a conventional rhombic unit cell with two sites. In specific, we assign $J_{ij}=J_0$ inside unit cells and $J_{ij}=J_1$ between unit cells [see Fig.~\ref{Fig:1}], which respects the $C_{6v}$ symmetry. The six-site cluster in a unit cell can be regarded as an ``artificial molecule''.

The spin-wave excitations in Hamiltonian (1) can be represented as magnons by the Holstein-Primakoff transformation \cite{Holstein1940},
\begin{subequations}
\begin{eqnarray}
S_i^z&=&S_i-b_i^\dag b_i,\\S_i^+&=&\biggl(\sqrt{2S_i-b_i^\dag b_i}\biggr)b_i,\\S_i^-&=&b_i^\dag\sqrt{2S_i-b_i^\dag b_i},
\label{eqHPT} % Holstein-Primakoff Transformation
\end{eqnarray}
\end{subequations}
with $S_i^\pm=S_i^x \pm iS_i^y$, where $S_i^x$, $S_i^y$ and $S_i^z$ are the three orthogonal components of $\bm{S}_i$, $b_i$ and $b_i^\dag$ are annihilation operators and creation operators of magnons. The Hamiltonian in the magnon representation is complicated since it contains many-body magnon-magnon interaction terms such as $b_i^\dag b_ib_j^\dag b_j$. At low temperatures, the average number of magnon excitations $\langle b_i^\dag b_i\rangle$ is small comparing to $2S_i$, making it a reasonable approximation to neglect the magnon-magnon interactions. Then the Holstein-Primakoff transformation gives $S_i^+\approx \sqrt{2S_i}b_i$ and $S_i^-\approx \sqrt{2S_i}b_i^\dag$, which results in the effective Hamiltonian with terms quadratic in $b_i$ and $b^\dagger_i$.

We then turn to momentum space by applying the Fourier transformation
\begin{eqnarray}
b_{i,\boldsymbol{r}}=\frac{1}{\sqrt{N}}\sum_{\bm{k}} b_{i,\bm{k}} {\rm{e}}^{{\rm{i}} \bm{k}\cdot \bm{r}},
\label{eqFT} % Fourier Transformation
\end{eqnarray}
where $i$ is the site index inside a unit cell as shown in Fig. \ref{Fig:1}, $N$ is the number of unit cells and $\bm{r}$ is the position of the unit cell. The Hamiltonian for magnons is then described as
\begin{eqnarray}
\hat{H}_{\rm{F}}=\sum_{\bm{k}}\Psi_{\bm{k}}^\dag H_{{\rm{F}},\bm{k}} \Psi_{\bm{k}},
\label{eqHfF} % Hamiltonian for Ferromagnets
\end{eqnarray}
where $\Psi_{\bm{k}}=[b_{1,\bm{k}}, b_{2,\bm{k}}, b_{3,\bm{k}}, b_{4,\bm{k}}, b_{5,\bm{k}}, b_{6,\bm{k}}]^T$ [see Fig. \ref{Fig:1}]. With the uniform spin length $S_i=S$, $H_{{\rm{F}},\bm{k}}$ is given by
\begin{equation}
H_{{\rm{F}},\bm{k}}=\left[\begin{array}{cc}
E_0I_3 & -Q_{\bm{k}}\\
-Q_{\bm{k}}^\dag & E_0I_3
\end{array} \right],
\label{eqHaM} % Hamiltonian Matrix
\end{equation}
with
 \begin{equation}
  Q_{\bm{k}}= S\left[\begin{array}{ccc}
J_1 {\rm{e}}^{-{\rm{i}} \bm{k}\cdot(\bm{a}_1-\bm{a}_2)} & J_0 & J_0 \\
J_0 & J_1 {\rm{e}}^{{\rm{i}} \bm{k}\cdot\bm{a}_1} & J_0 \\
J_0 & J_0 & J_1 {\rm{e}}^{-{\rm{i}} \bm{k}\cdot\bm{a}_2}
 \end{array}\right],
 \end{equation}
where $E_0=S(2J_0+J_1)$ and $I_3$ is a $3\times3$ identity matrix, and $\bm{a}_1$ and $\bm{a}_2$ are unit vectors presented in Fig. \ref{Fig:1}.

The eigenvalue equation with Hamiltonian (\ref{eqHaM}),
\begin{equation}
  H_{{\rm{F}},\bm{k}}\psi_n=\epsilon_{{\rm{F}},n,\bm{k}}\psi_n,
\label{EVE}
\end{equation}
gives the eigenstates and eigenvalues with $n=$1, 2, ..., 6. Then, Hamiltonian (\ref{eqHfF}) can be cast into
\begin{eqnarray}
\hat{H}_F=\sum_{\bm{k}}\sum_n \epsilon_{{\rm{F}},n,\bm{k}} b_{n,\bm{k}}'^{\dag} b_{n,\bm{k}}',
\end{eqnarray}
where $b_{n,\bm{k}}^{'\dag}=\sum_i\psi_{ni}b_{i,\bm{k}}^{\dag}$ is the creation operator of $n$-th eigen magnon mode and $\psi_{ni}$ is the $i$-th element of $\psi_n$.

We notice that $H_{{\rm{F}},\bm{k}}$ in Eq. (\ref{eqHaM}) only differs from the electronic model on honeycomb lattice with hopping textures \cite{WuLH2016,Kariyado2017},
\begin{equation}
H_{\bm{k}}=\left[\begin{array}{cc}
0 & Q_{\bm{k}}\\
Q_{\bm{k}}^\dag & 0
\end{array} \right],
\label{HoT1T0}
\end{equation} % Hamiltonian of t1t0 model
by the diagonal element $E_0$, where $t_0$ and $t_1$ in electronic model are replaced by $SJ_0$ and $SJ_1$ respectively. Therefore, they share the same eigenstates with the shifted eigenvalues:
\begin{equation}
\epsilon_{{\rm{F}},n,\bm{k}}=E_0-\epsilon_{0,n,\bm{k}},
\label{EoF} % energy of Ferromagnets
\end{equation}
where $\epsilon_{0,n,\bm{k}}$ is the eigenvalue of $H_{\bm{k}}$ in Eq. (\ref{HoT1T0}). One has $\epsilon_{{\rm{F}},n,\bm{k}}\in[0, 2E_0]$, since $\epsilon_{0,n,\bm{k}}\in[-E_0, E_0]$.

The time-dependent form of  $b_{n,\bm{k}}^{'}$ is $b_{n,\bm{k}}^{'}{\rm{e}}^{-{\rm{i}} \omega_{{\rm{F}},n,\bm{k}} t}$, where $\omega_{{\rm{F}},n,\bm{k}}$ is the magnon frequency. With the Heisenberg equation of motion
\begin{eqnarray}
\frac{d b_{n,\bm{k}}^{'}}{d t}=\frac{{\rm{i}}}{\hbar}[\hat{H}_F, b_{n,\bm{k}}^{'}]
\end{eqnarray}
and the commutation relation of bosons
\begin{eqnarray}
\left[b_{n,\bm{k}}, b^{\dag}_{n',\bm{k}'}\right]=\delta_{n n'} \delta_{\bm{k} \bm{k}'},
\label{eqCRB}  % Commutation Relation of Bosons
\end{eqnarray}
 we can derive the dynamic equation of magnon $b_{n,\bm{k}}^{'}$
\begin{eqnarray}
\frac{d^2 b_{n,\bm{k}}^{'}}{d t^2}=-\frac{\epsilon_{{\rm{F}},n,\bm{k}}^2}{\hbar^2} b_{n,\bm{k}}^{'}.
\end{eqnarray}
The magnon frequency can be obtained by
\begin{equation}
\omega_{{\rm{F}},n,\bm{k}}=\epsilon_{{\rm{F}},n,\bm{k}}/\hbar,
\label{EOFaE} % energy of Ferromagnets
\end{equation}
since $\epsilon_{{\rm{F}},n,\bm{k}}>0$.

\begin{figure}[!htb]
\centering
\includegraphics[clip=true,width=0.6\textwidth]{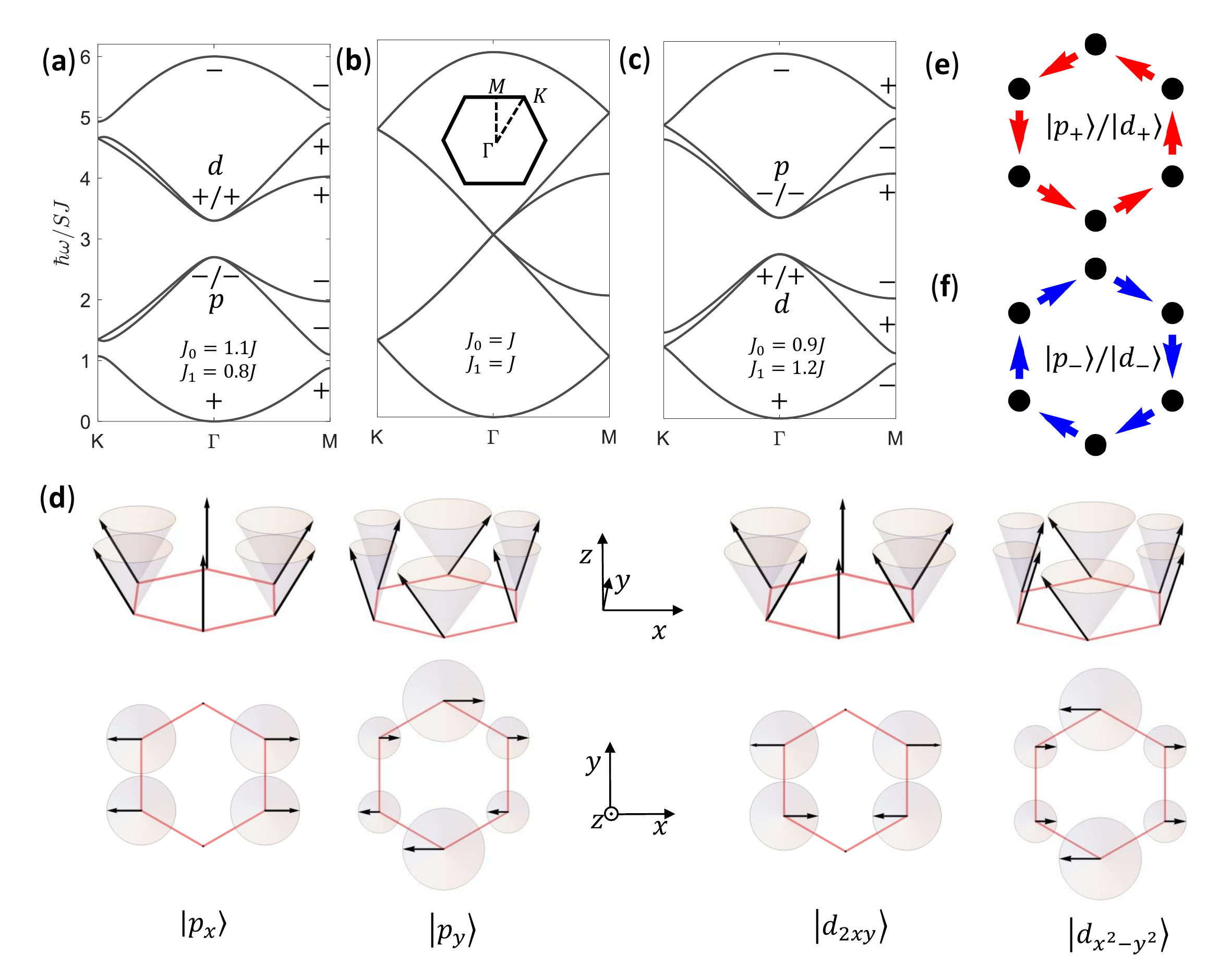}
\caption{(a) Frequency band structure of magnon modes for a ferromagnet with $J_0>J_1$, where a gap opens between $d$ and $p$ modes. The parity of inversion symmetry at the M point is the same as that at the $\Gamma$ point for each band, where the even/odd parity is marked by $+$/$-$.  (b) Same as (a) except for $J_0=J_1$, where the gap closes and Dirac cones appear. (c) Same as (a) except for $J_0<J_1$, where the gap reopens and a band inversion of $p$ and $d$ modes at the $\Gamma$ point takes place, and below the gap the number of eigenstates with even parity at the $\Gamma$ point does not equal to that at the M point. For simplicity, couplings are chosen to satisfy $2J_0+J_1=3J$ and $|J_0-J_1|=0.3J$, to make gaps in (a) and (c) overlapped. (d) Magnon modes at the $\Gamma$ point, with the spin precession denoted by gray cones, where the projections of spins on $xy$ plane satisfy the parity of $p$ and $d$ basis functions respectively. (e) and (f) Schematic magnon currents of magnon modes with up- and down-pseudospin, circulating counterclockwise and clockwise respectively, where $|p_{\pm}\rangle=\frac{1}{\sqrt{2}}(|p_x\rangle\pm {\rm{i}}|p_{y}\rangle)$ and $|d_{\pm}\rangle=\frac{1}{\sqrt{2}}(|d_{x^2-y^2}\rangle\pm {\rm{i}}|d_{2xy}\rangle)$.}
\label{Fig:2}
\end{figure}

The frequency band structures of magnon modes are shown in Fig. \ref{Fig:2}. For $J_0=J_1$, the frequency band structure exhibits Dirac cones same as that of the electronic structure of graphene \cite{Fransson2016}. These Dirac cones are folded to the $\Gamma$ point as shown in Fig. \ref{Fig:2}(b), since we are using the six-site hexagonal unit cell instead of the conventional two-site rhombic unit cell. For $J_0\neq J_1$, a frequency gap opens at the $\Gamma$ point as displayed in Figs. \ref{Fig:2}(a) and \ref{Fig:2}(c). As clarified in Ref. 32, the band structures for $J_0>J_1$ [Fig.~\ref{Fig:2}(a)] and for $J_0<J_1$ [Fig.~\ref{Fig:2}(c)] are distinct in topology.

Because Hamiltonian (\ref{eqHM}) preserves the inversion symmetry, eigen wavefunctions of Hamiltonian (\ref{eqHaM}) can be indexed by parity eigenvalues, even and odd, at the $\Gamma$ point and M point. In the case $J_0>J_1$ [see Fig. \ref{Fig:2}(a)], the two degenerate wavefunctions below the gap are $|p_x\rangle$ and $|p_y\rangle$ with odd parity at the $\Gamma$ point, whereas the two degenerate wavefunctions above the gap are $|d_{x^2-y^2}\rangle$ and $|d_{2xy}\rangle$ with even parity. The $p$ and $d$ magnon modes are shown in Figs. \ref{Fig:2}(d). The parity of each band at the M point is the same as that at the $\Gamma$ point, denoting a topologically trivial state \cite{Benalcazar2014,Kariyado2018}. In the case $J_0<J_1$, the frequencies of $p$ and $d$ modes are inverted at the $\Gamma$ point as shown in Fig. \ref{Fig:2}(c). Now, at the $\Gamma$ point, all three eigenstates below the gap are of even parity, while at the M point, there are two eigenstates with odd parity and one eigenstate with even parity. The unequal numbers of eigenstates with even parity indicate a topological state \cite{Benalcazar2014,Kariyado2018}. The band structure of topological state in Fig. \ref{Fig:2}(c) cannot be continually transformed from that of the topologically trivial state in Fig. \ref{Fig:2}(a) without closing the gap [see Fig. \ref{Fig:2}(b)].

The double degeneracy of $p$ ($d$) modes at the $\Gamma$ point are protected by the $C_{6v}$ symmetry, where the wavefunctions of $p$ ($d$) modes correspond to the basis functions of two-dimensional irreducible representation $E_1$ ($E_2$). In order to better understand the symmetry at the $\Gamma$ point, we introduce a pseudo time-reversal operator $\mathcal{T}={\rm{i}}\sigma_y \mathcal{K}$, where $\sigma_y$ is Pauli matrix referring to the basis functions of $E_1$ and $E_2$, and $\mathcal{K}$ denotes the complex conjugate operator \cite{WLH2015,WuLH2016}. This pseudo time-reversal operator originates from the $C_{6v}$ symmetry and the time-reversal symmetry of the system, which produces the Kramers doubling in the present bosonic system. Kramers pairs associated with $p$ or $d$ orbitals are constructed as $|p_{\pm}\rangle=\frac{1}{\sqrt{2}}(|p_x\rangle\pm {\rm{i}}|p_{y}\rangle)$ and $|d_{\pm}\rangle=\frac{1}{\sqrt{2}}(|d_{x^2-y^2}\rangle\pm {\rm{i}}|d_{2xy}\rangle)$ respectively, which carry specific pseudospins defined on the ``artificial molecule''. In the present system, the magnon current $I_{ij}=\langle \hat{I}_{ij}\rangle$ is given by the current operator
\begin{eqnarray}
\hat{I}_{ij}=-\frac{{\rm{i}} SJ_{ij}}{\hbar}(b_i^\dag b_j-b_j^\dag b_i).
\end{eqnarray}
For magnon modes with pseudospin up/down, magnon currents circulate counterclockwise/clockwise, as shown in Figs. \ref{Fig:2}(e) and \ref{Fig:2}(f).

\begin{figure}[tb]
\centering
  \includegraphics[clip=true,width=0.7\textwidth]{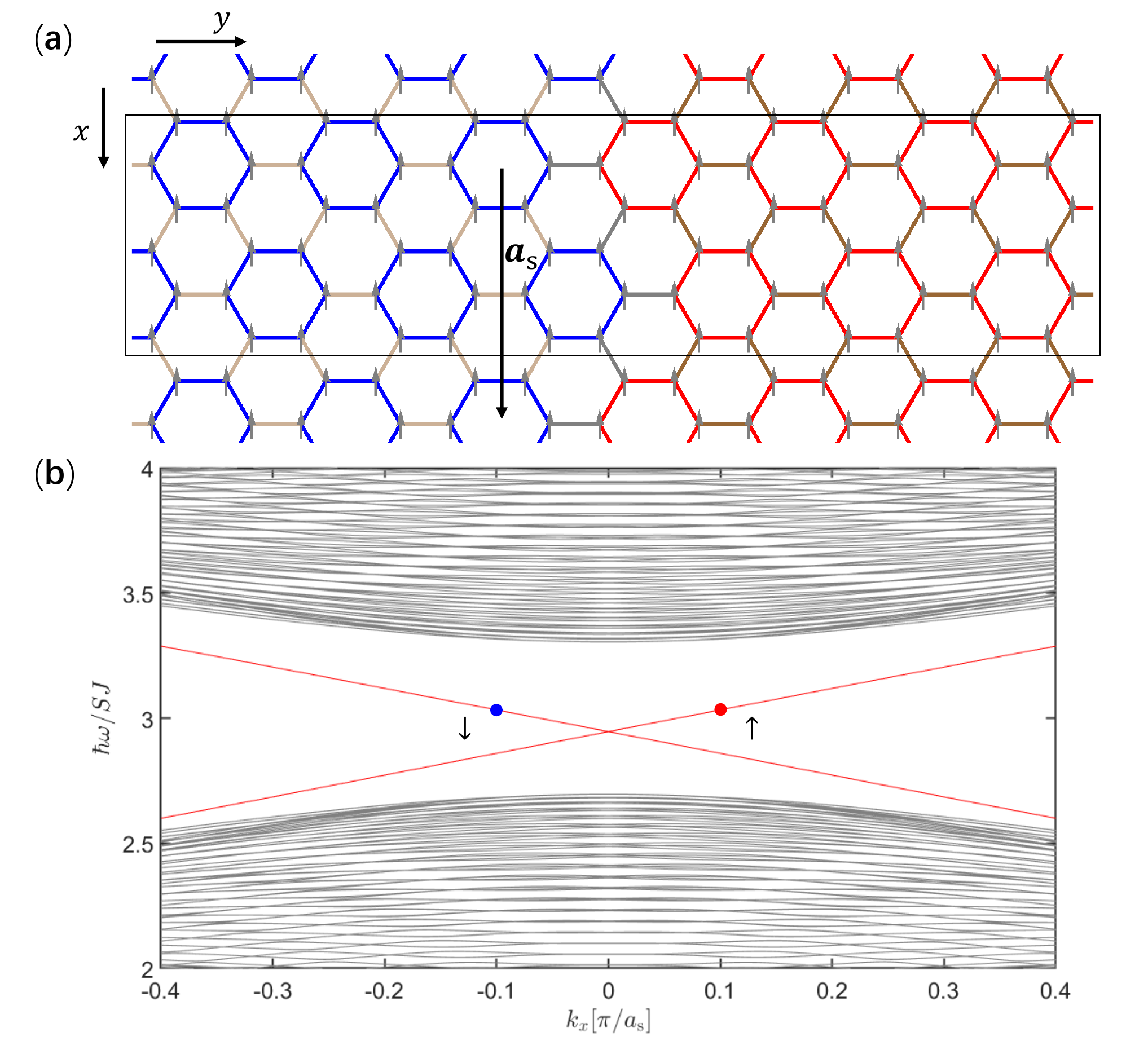}
  \caption{(a) Ferromagnetic heterostructure including a trivial domain and a topological domain, which is uniform and infinitely long in the $x$ direction. In the $y$ direction, 60 unit cells are contained in both trivial and topological domains periodically, where the exchange couplings are the same as in Figs. \ref{Fig:2}(a) and \ref{Fig:2}(c) respectively. The exchange couplings inside unit cells are denoted by blue and red lines, whereas those between unit cells are denoted by light and dark brown lines in the trivial domain and the topological domain respectively. The exchange couplings between the trivial and topological domains (denoted by gray lines) are given by the geometric mean of value of $J_1$ in Figs. \ref{Fig:2}(a) and \ref{Fig:2}(c). (b) Frequency band structure of magnon modes calculated based on the supercell denoted by the rectangular frame with $\bm{a}_{\rm{s}}$ the unit vector, where topological interface dispersions (red lines) appear in the bulk band gap.}
  \label{Fig:3}
\end{figure}

Around the $\Gamma$ point, magnon modes are predominantly occupied by the $p$ orbitals and $d$ orbitals. We can rewrite Hamiltonian (\ref{eqHaM}) in the basis of $[|p_+\rangle, |d_+\rangle, |p_-\rangle, |d_-\rangle]$ up to the lowest order of momentum $\bm{k}$ as
\begin{equation}
  H_{{\rm{eff}},\bm{k}}=\left[\begin{array}{cc}
 H_{{\rm{pd}},\bm{k}}& 0\\
0 & H^*_{{\rm{pd}},-\bm{k}}
 \end{array}\right]
 \label{EHoPD}
 \end{equation}
with
\begin{equation}
 H_{{\rm{pd}},\bm{k}}= E_0\left[\begin{array}{cc}
 1 & 0 \\
0 & 1
 \end{array}\right]+
 S\left[\begin{array}{cc} (J_1-J_0)-\frac{9}{2}J_1 a_0^2\bm{k}^2 & -\frac{3}{2}{\rm{i}} J_1 a_0(k_x+{\rm{i}} k_y) \\
\frac{3}{2}{\rm{i}} J_1 a_0(k_x-{\rm{i}} k_y) & (J_0-J_1)+\frac{9}{2}J_1 a_0^2\bm{k}^2
 \end{array}\right],
 \label{EHoPD2}
 \end{equation}
where $a_0$ is the length of unit vector [see Fig.\ref{Fig:1}]. This effective Hamiltonian is similar to the Bernevig-Hughes-Zhang model for quantum spin Hall effect and that found for topological photonic crystals \cite{Bernevig2006,WLH2015}. In Hamiltonian (\ref{EHoPD2}), $E_0$ simply makes a constant shift in the eigenvalues, and
\begin{equation}
 M=J_0-J_1
 \label{EDM}
 \end{equation}
 is the effective Dirac mass, which is positive in a topologically trivial state. In the case  $J_0<J_1$, M becomes negative and a band inversion takes place, which turns the system into a topological state with a topological gap $2(J_1-J_0)$, as discussed above and shown in Fig. \ref{Fig:2}(c). It is noticed that the topological magnon modes in the present system relies on crystalline symmetry and has a weaker topology comparing to those systems with the DM interaction. Nevertheless, we expect that stable unidirectional interface states between topological and trivial domains can be observed experimentally, similarly to photonic systems \cite{HeC2016,LiY2018,Yang2018,Shao2020}.

\begin{figure}[tb]
\centering
  \includegraphics[clip=true,width=0.6\textwidth]{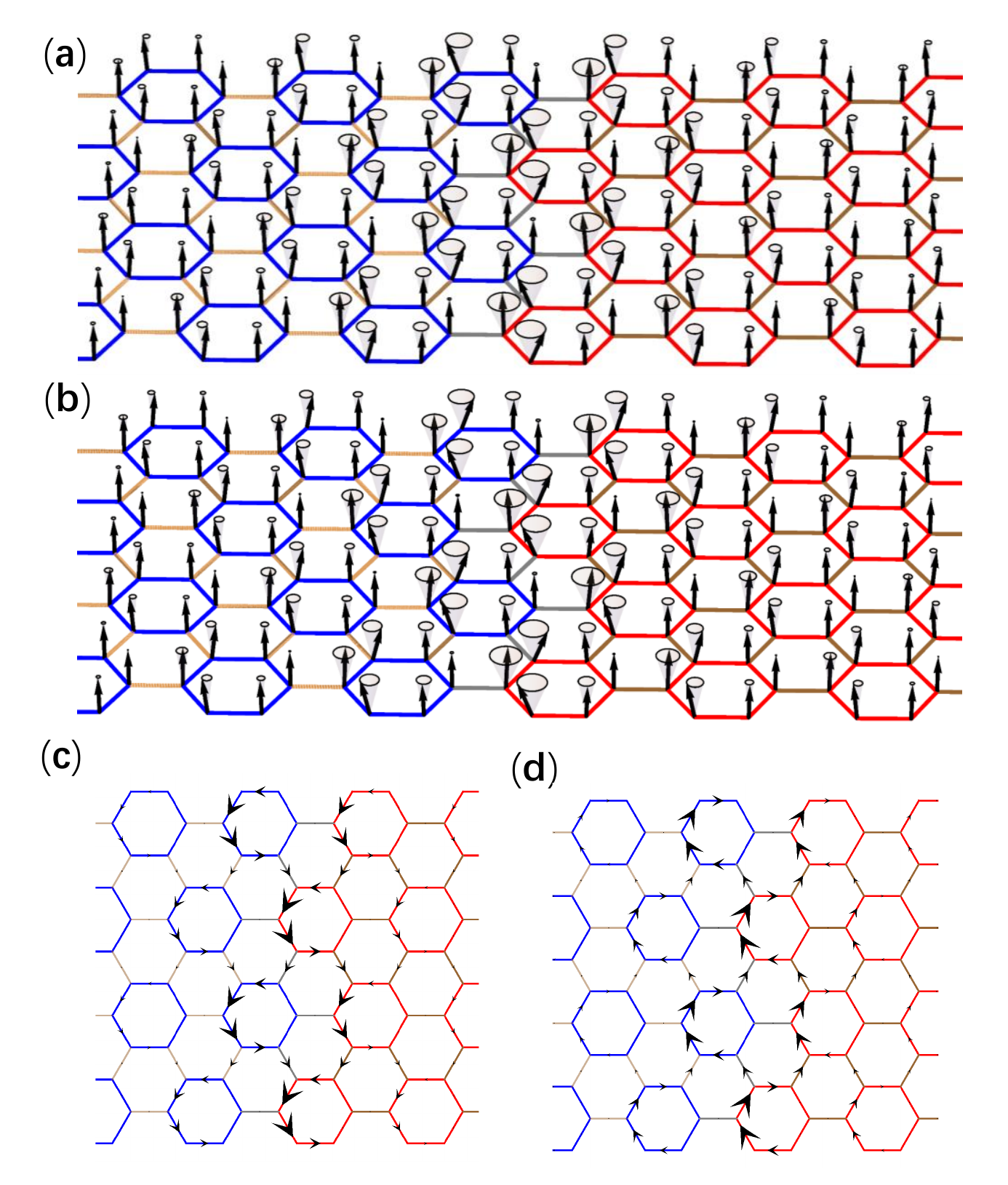}
  \caption{ (a) and (b) Distribution of magnon modes at the two opposite momenta denoted by $\uparrow$ and $\downarrow$ in Fig. \ref{Fig:3}(b) respectively. (c) and (d) Same as (a) and (b) except for distributions of magnon currents between sites presented by black arrows. The magnon mode denoted by $\uparrow$/$\downarrow$ carries up-/down-pseudospin, showing counterclockwise/clockwise magnon currents in unit cells.}
  \label{Fig:4}
\end{figure}

In order to investigate topological interface magnon modes, we consider a heterostructure as illustrated in Fig. \ref{Fig:3}(a), which is uniform and infinitely long in the $x$ direction. In the $y$ direction, the heterostructure contains 60 unit cells in both trivial and topological domains periodically, for the simplicity of calculation. For each interface, two interface dispersions appear inside the bulk gap in the frequency band structure shown in Fig. \ref{Fig:3}(b), which is obtained numerically using the large unit cell for the heterostructure [Fig.~\ref{Fig:3}(a)]. The interface magnon modes at the two momenta marked by dots in Fig. \ref{Fig:3}(b) are shown in Figs. \ref{Fig:4}(a) and \ref{Fig:4}(b) respectively. The topological magnon modes correspond to precessions of spins around the $z$ axis, with amplitudes decaying into the bulks exponentially. In Figs. \ref{Fig:4}(c) and \ref{Fig:4}(d), magnon currents of topological interface magnon modes with counterclockwise circulation and clockwise circulation in unit cells correspond to the up- and down-pseudospin respectively. The net current for up-/down-pseudospin flows to the positive/negative $x$ direction, manifesting the pseudospin-momentum locking in the present topological magnon modes.

\begin{figure}[t]
\centering
  \includegraphics[clip=true,width=0.6\textwidth]{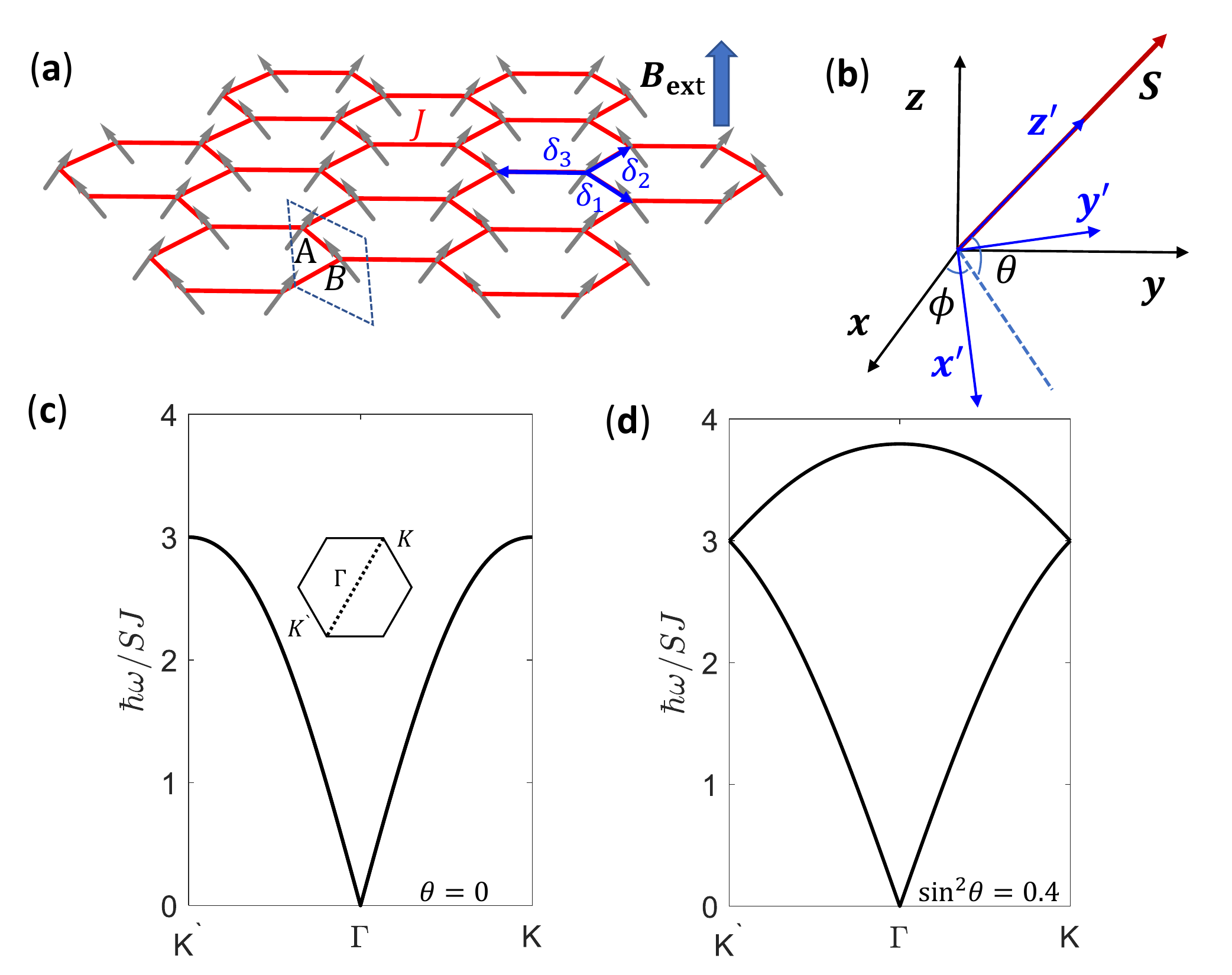}
  \caption{(a) Antiferromagnet on honeycomb lattice under an external magnetic field $B_{\rm{ext}}$ perpendicular to the lattice. The rhombic unit cell denoted by the dashed line contains two sites A and B. (b) Transformation between the laboratory frame and the rotating frame. $x$, $y$ and $z$ (denoted by black arrows) are coordinate axes of the laboratory frame, where the external magnetic field $B_{\rm{ext}}$ induces a canting angle $\theta$ of spin relative to the $xy$ plane, while $x'$, $y'$ and $z'$ (denoted by blue arrows) are coordinate axes of the rotating frame. The transformation is achieved by rotating coordinate axes around the $z$ axis with an angle $\phi$ then around the $y'$ axis with an angle $\pi/2-\theta$. (c) Frequency band structure of magnon modes with double degenerate dispersions for an antiferromagnet in absence of external magnetic field. (d) Frequency band structure of magnon modes for a canted antiferromagnet induced by an external magnetic field corresponding to ${\rm{sin}}^2\theta=0.4$ (see text for details), where the degeneracy in (c) is lifted except for the $K$ and $K'$ points.}
  \label{Fig:5}
\end{figure}

\subsection*{Topological magnon modes in antiferromagnets}

Next, we explore topological magnon modes in antiferromagnets on honeycomb lattice. Antiferromagnets exhibit degenerate dispersions guaranteed by the combination of time-reversal and inversion symmetries, where no Dirac cone exists \cite{Owerre2017_3}. Therefore, it is not straightforward to realize topological magnon modes using the procedure formulated for ferromagnets directly. In order to overcome this difficulty, we introduce an external magnetic field to lift the double degeneracy. The Heisenberg model for antiferromagnets is
\begin{eqnarray}
  H=\sum_{\langle i,j\rangle}J_{ij}\bm{S}_i\cdot \bm{S}_j-B_{\rm{ext}}\sum_iS^z_i,
  \label{eqHMfAF}  % Heisenberg Model for AntiFerromagnets
\end{eqnarray}
where exchange couplings $J_{ij}>0$, the external magnetic field $B_{\rm{ext}}>0$, and spin lengths $S_i=S$, and for simplicity, the external magnetic field is perpendicular to the two-dimensional spin plane. Without single ion anisotropy considered here, spins are automatically aligned in the plane perpendicular to the direction of external magnetic field and canted as shown schematically in Fig. \ref{Fig:5}(a). The canting angle $\theta$ of an antiferromagnet under external magnetic field is determined by the competition between exchange couplings and the Zeeman energy. We consider uniform exchange couplings $J_{ij}=J$ as shown in Fig. \ref{Fig:5}(a). The energy of a classical canted antiferromagnetic state with $N$ rhombic unit cells is
\begin{eqnarray}
E_{\rm{classical}}=-3JNS^2{\rm{cos}}2\theta-2B_{\rm{ext}}NS{\rm{sin}}\theta,\label{eq:Eclassical_AF}
\end{eqnarray}
which is minimized when spins are canted at an angle ${\rm{sin}}\theta=B_{\rm{ext}}/(6JS)$.

In order to apply the Holstein-Primakoff transformation for antiferromagnets, we need a rotating frame where the $z$ coordinate axis is rotated to the local spin direction at each site [see Fig. \ref{Fig:5}(b)]. The transformation between the rotating frame and the laboratory frame is
\begin{equation}
  \left[\begin{array}{c}S^x\\S^y\\S^z\end{array}\right]= \left[\begin{array}{ccc}
  {\rm{sin}}\theta{\rm{cos}}\phi & -{\rm{sin}}\phi & {\rm{cos}}\theta{\rm{cos}}\phi \\
{\rm{sin}}\theta{\rm{sin}}\phi & {\rm{cos}}\phi & {\rm{cos}}\theta{\rm{sin}}\phi \\
-{\rm{cos}}\theta & 0 & {\rm{sin}}\theta
 \end{array}\right]\left[\begin{array}{c}S'^{x}\\S'^{y}\\S'^{z}\end{array}\right],
 \label{FRT}
 \end{equation}
where $S^x$, $S^y$, $S^z$ are spin components of the laboratory frame, $S'^{x}$, $S'^{y}$, $S'^{z}$ are those of the rotating frame, $\phi$ is the angle between the projection of a spin in the $xy$ plane and the coordinate axis $x$ and $\theta$ is the canting angle of spin. Note that projections of spins on different sublattices point to opposite directions on the $xy$ plane, namely $\phi=\phi_0$ ($\phi=\pi+\phi_0$) for spins on A (B) sites, a classical N\'eel order to host magnon excitations. Now Hamiltonian (\ref{eqHMfAF}) becomes
\begin{eqnarray}
\label{eqAFRF}
H&&=J\sum_{\langle i,j\rangle}[{\rm{cos}}2\theta(S'^{x}_iS'^{x}_j-S'^{z}_iS'^{z}_j)-S'^{y}_iS'^{y}_j-{\rm{sin}}2\theta(S'^{x}_iS'^{z}_j+S'^{z}_iS'^{x}_j)]\\&&\nonumber-B\sum_i[{\rm{sin}}\theta S'^{z}_i-{\rm{cos}}\theta S'^{x}_i],
\end{eqnarray}  % AntiFerromagnets in Rotation Frame
which does not depend explicitly on $\phi_0$. Adapting the Holstein-Primakoff transformation (2) for $S'^{x,y,z}_i$ and applying the Fourier transformation (\ref{eqFT}), a Bogoliubov Hamiltonian for magnon excitations is derived from Hamiltonian (\ref{eqAFRF}) as
\begin{eqnarray}
H=\frac{1}{2}\sum_{\bm{k}}\Psi_{\bm{k}}^\dag H_{{\rm{CAF}},\bm{k}} \Psi_{\bm{k}},
\label{eqHfAF}  % Hamiltonian for Antiferromagnets
\end{eqnarray}
where
\begin{equation}
H_{{\rm{CAF}},\bm{k}}=JS\left[\begin{array}{cccc}
3 & -{\rm{sin}}^2\theta f_{\bm{k}} & 0 & (1-{\rm{sin}}^2\theta)f_{\bm{k}} \\
-{\rm{sin}}^2\theta f_{\bm{k}}^* & 3 & (1-{\rm{sin}}^2\theta)f_{\bm{k}}^* & 0 \\
0 & (1-{\rm{sin}}^2\theta)f_{\bm{k}} & 3 & -{\rm{sin}}^2\theta f_{\bm{k}} \\
(1-{\rm{sin}}^2\theta)f_{\bm{k}}^* & 0 & -{\rm{sin}}^2\theta f_{\bm{k}}^* & 3
 \end{array}\right],
 \label{HoCAFoR}
 \end{equation}
$\Psi_{\bm{k}}=[b_{A,\bm{k}}, b_{B,\bm{k}}, b^\dag_{A,-\bm{k}}, b^\dag_{B,-\bm{k}}]^T$, $f_{\bm{k}}=\sum_i {\rm{e}}^{{\rm{i}}\bm{k}\cdot\delta_i}$ with $\delta_i$ being the n.n. vectors shown in Fig. \ref{Fig:5}(a). Note that the terms linear in $b_i$ or $b^\dagger_i$ cancel out for ${\rm{sin}}\theta=B_{\rm{ext}}/(6JS)$, which minimizes Eq.~\eqref{eq:Eclassical_AF}. Because the classical N\'eel order is not the ground state for quantum antiferromagnets, $bb$ and $b^\dag b^\dag$ terms that do not conserve the total number of magnon appear, as such the basis vector $\Psi_{\bm{k}}$ including both creation and annihilation operators is chosen.

Dispersions of magnon modes can be derived by applying the bosonic Bogoliubov transformation \cite{COLPA1978327,vanHemmen1980} $\Psi_{\bm{k}}=T_{\bm{k}}\Gamma_{\bm{k}}$ to diagonalize $H_{{\rm{CAF}},\bm{k}}$,
\begin{eqnarray}
\Psi_{\bm{k}}^\dag H_{{\rm{CAF}},\bm{k}} \Psi_{\bm{k}}=\Gamma_{\bm{k}}^\dag E_{\bm{k}} \Gamma_{\bm{k}},
\label{BBT} % Bosonic Bogoliubov Transformation
\end{eqnarray}
where $\Gamma_{\bm{k}}=[\gamma_{1,\bm{k}}, \gamma_{2,\bm{k}}, \gamma^\dag_{1,-\bm{k}}, \gamma^\dag_{2,-\bm{k}}]^T$ is a new set of creation and annihilation operators for magnons and $E_{\bm{k}}$ is a diagonalized matrix. Magnons both before and after the Bogoliubov transformation should obey the same commutation relation for bosons in Eq. (\ref{eqCRB}), demanding that the matrix $T_{\bm{k}}$ satisfies $T_{\bm{k}}\hat{I}T^\dag_{\bm{k}}=\hat{I}$ with $\hat{I}=\rm{diag}(1, 1, -1, -1)$. Here, $T_{\bm{k}}$ is not a unitary matrix since $(T^\dag_{\bm{k}})^{-1}=\hat{I}T_{\bm{k}}\hat{I}$, which leaves $H_{{\rm{CAF}},\bm{k}}T_{\bm{k}}=(T^\dag_{\bm{k}})^{-1}E_{\bm{k}} \neq T_{\bm{k}}E_{\bm{k}}$ in general. Therefore, one cannot calculate {$E_{\bm{k}}$ from the characteristic equation of matrix $H_{{\rm{CAF}},\bm{k}}$ directly, since columns of $T_{\bm{k}}$ are not eigenvectors of $H_{{\rm{CAF}},\bm{k}}$.

Multiplying $\hat{I}(T^\dag_{\bm{k}})^{-1}$ to both sides of $T^\dag_{\bm{k}}H_{{\rm{CAF}},\bm{k}}T_{\bm{k}}=E_{\bm{k}}$ from left, we obtain the relation
\begin{equation}
\hat{I}H_{{\rm{CAF}},\bm{k}}T_{\bm{k}}=T_{\bm{k}}\hat{I}E_{\bm{k}},
\label{EVEfCAF}
\end{equation}
namely columns of $T_{\bm{k}}$ are eigenvectors of $\hat{I}H_{{\rm{CAF}},\bm{k}}$. The diagonalized matrix $\hat{I}E_{\bm{k}}$ can be derived from the characteristic equation of matrix $\hat{I}H_{{\rm{CAF}},\bm{k}}$, ${\rm{det}}|\hat{I}H_{{\rm{CAF}},\bm{k}}-\hat{I}E_{\bm{k}}|=0$, which is the conventional way to obtain $\hat{I}E_{\bm{k}}$ and $T_{\bm{k}}$.

Because $\Psi_{\bm{k}}$ and $\Gamma_{\bm{k}}$ include both creation and annihilation operators of magnons, dispersions in Eq. (\ref{BBT}) are redundant, which should be eliminated by symmetries. First, we define a particle-hole operator
\begin{eqnarray}
\mathcal{C}=\left[\begin{array}{cc}
0 & I_2 \\
I_2 & 0
\end{array}\right]\mathcal{KP},
\label{PHO}
\end{eqnarray}
where $I_2$ is a $2\times2$ identity matrix, $\mathcal{K}$ is the complex conjugate operator and $\mathcal{P}$ is the space-inversion operator. For $\hat{I}H_{{\rm{CAF}},\bm{k}}$, we can directly check from Eq. (\ref{HoCAFoR}) and Eq. (\ref{PHO}) that $\hat{I}H_{{\rm{CAF}},\bm{k}}\mathcal{C}=-\mathcal{C}\hat{I}H_{{\rm{CAF}},\bm{k}}$, leading to $\hat{I}E_{\bm{k}}={\rm{diag}}(\epsilon_{1, \bm{k}}, \epsilon_{2, \bm{k}}, -\epsilon_{1, \bm{k}}, -\epsilon_{2, \bm{k}})$. With the same method, we can also check that $\mathcal{P}\hat{I}H_{{\rm{CAF}},\bm{k}}=\hat{I}H_{{\rm{CAF}},-\bm{k}}\mathcal{P}$, which leads to $\epsilon_{n, \bm{k}}=\epsilon_{n, -\bm{k}}$ ($n=1,2$). In this way, Hamiltonian (\ref{eqHfAF}) is diagonalized into \cite{COLPA1978327,vanHemmen1980,Watabe1995}
\begin{eqnarray}
\hat{H}_{\rm{CAF}}=\sum_{n=1,2;\bm{k}}\epsilon_{n,\bm{k}}\gamma^\dag_{n,\bm{k}}\gamma_{n,\bm{k}},
\label{EVoAT}
\end{eqnarray}
with
\begin{equation}
\gamma^\dag_{n,\bm{k}}=\sum_i(u_{i,n,\bm{k}}b^\dag_{i,\bm{k}}-v^*_{i,n,-\bm{k}}b_{i,-\bm{k}}),
\label{ESoCA}
\end{equation}
where $(u_{{\rm{A}},n,\bm{k}},u_{{\rm{B}},n,\bm{k}},v^*_{{\rm{A}},n,-\bm{k}},v^*_{{\rm{B}},n,-\bm{k}})$ is the $n$-th column of $T_{\bm{k}}$.

\begin{figure}[t]
\centering
  \includegraphics[clip=true,width=0.9\textwidth]{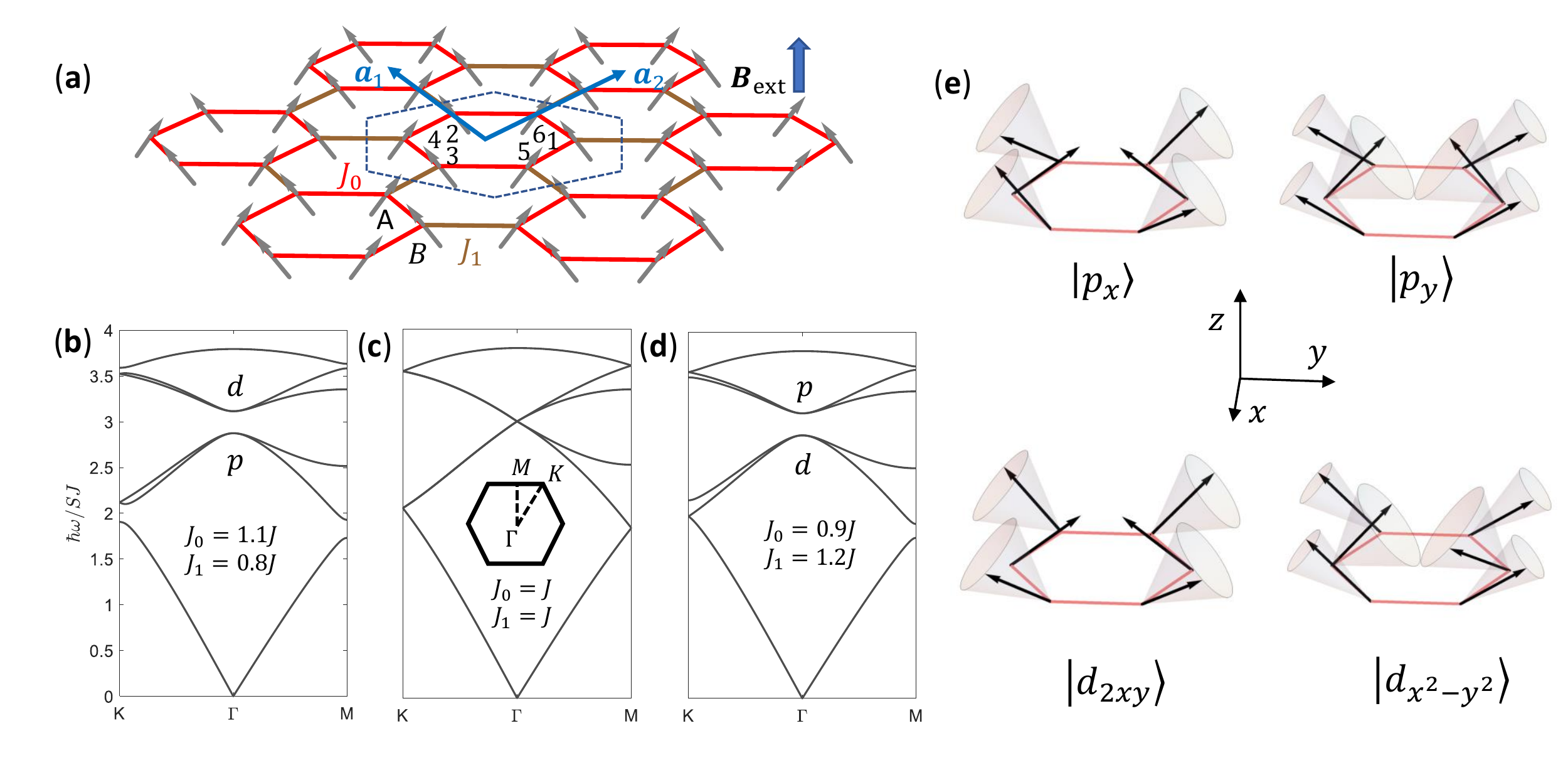}
  \caption{(a) Same as Fig. \ref{Fig:5}(a) except for that a coupling texture is introduced, where hexagonal unit cells are chosen with exchange couplings $J_0$/$J_1$ inside/between unit cells. (b) Frequency band structure of magnon modes for a canted antiferromagnet with $J_0>J_1$. (c) Same as (b) except for $J_0=J_1$. The $K$ and $K^{'}$ points in Fig. \ref{Fig:5}(d) are folded into the $\Gamma$ point. (d) Same as (b) except for $J_0<J_1$. Canting angle ${\rm{sin}}^2\theta=0.4$ is considered for (b), (c) and (d). (e) Magnon modes for canted antiferromagnets at the $\Gamma$ point. In the rotating frame defined in Fig. \ref{Fig:5}(a), where the rotating frame at A and B sublattice are different by a $\pi$ rotation of $\phi$, magnon modes look exactly the same way as Fig. \ref{Fig:2}(d). For simplicity, $\phi=\pm\pi/2$ is taken.}
  \label{Fig:6}
\end{figure}

Then the dispersions of magnon modes for the canted antiferromagnet shown in Fig. \ref{Fig:5}(a) can be derived from the above procedure explicitly as \cite{Owerre2017_3}
 \begin{subequations}
\begin{eqnarray}
\epsilon_{1,\bm{k}}=SJ\sqrt{9+(2{\rm{sin}}^2\theta-1)|f_{\bm{k}}|^2 + 6{\rm{sin}}^2\theta|f_{\bm{k}}|},\\\epsilon_{2,\bm{k}}=SJ\sqrt{9+(2{\rm{sin}}^2\theta-1)|f_{\bm{k}}|^2 - 6{\rm{sin}}^2\theta|f_{\bm{k}}|}.
\end{eqnarray}
\end{subequations}
It is clear that without an external magnetic field $(\theta=0)$, the dispersion is doubly degenerate in the whole Brillouin zone [see Fig. \ref{Fig:5}(c)] where no Dirac cone exists, unlike ferromagnets [see Fig. \ref{Fig:2}(b)]. A finite external magnetic field lifts the degeneracy, except for the $K$ and $K'$ points where $f_K=f_{K'}=0$ and $\epsilon_{i,K}=\epsilon_{i,K'}=3SJ$, as shown in Fig. \ref{Fig:5}(d).

By choosing a hexagonal unit cell for canted antiferromagnets with a coupling texture as shown in Fig. \ref{Fig:6}(a), where the canting angle is determined by ${\rm{sin}}\theta=B_{\rm{ext}}/2S(2J_0+J_1)$ now, Hamiltonian (\ref{HoCAFoR}) becomes
\begin{equation}
H_{{\rm{CAF}},\bm{k}}=\left[\begin{array}{cc}
(1-{\rm{sin}}^2\theta)E_0I_6-{\rm{sin}}^2\theta H_{{\rm{F}},\bm{k}} &  (1-{\rm{sin}}^2\theta)(E_0I_6-H_{{\rm{F}},\bm{k}}) \\
(1-{\rm{sin}}^2\theta)(E_0I_6-H_{{\rm{F}},\bm{k}})  & (1-{\rm{sin}}^2\theta)E_0I_6-{\rm{sin}}^2\theta H_{{\rm{F}},\bm{k}}
 \end{array}\right],
 \label{HoCAFoH}
 \end{equation}
 where $E_0=S(2J_0+J_1)$, $I_6$ is a $6\times6$ identity matrix and $H_{{\rm{F}},\bm{k}}$ is Hamiltonian (\ref{eqHaM}) for ferromagnets.

Eigenvalues of canted antiferromagnets can be derived in terms of eigenvalues of ferromagnets with the same coupling texture. Here we consider the particle solution of Eq. (\ref{EVEfCAF}) which has positive energy $\epsilon_{n,\bm{k}}$ with $n=$1, 2, ..., 6. All four blocks of $H_{{\rm{CAF}},\bm{k}}$ in Eq. (\ref{HoCAFoH}) are digonalized in the basis of eigenstates of $H_{{\rm{F}},\bm{k}}$, so that for an eigenstate $\psi$ of $H_{{\rm{F}},\bm{k}}$ with eigenvalue $\epsilon_{{\rm{F}},n,\bm{k}}$, the state $(\psi, \alpha\psi)^T$ is an eigenstate of $\hat{I}H_{{\rm{CAF}},\bm{k}}$ in Eq. (\ref{EVEfCAF}), provided
\begin{subequations}
\begin{eqnarray}
(1-{\rm{sin}}^2\theta)(1+\alpha E_0)-({\rm{sin}}^2\theta+\alpha-\alpha{\rm{sin}}^2\theta)\epsilon_{{\rm{F}},n,\bm{k}}&=&\epsilon_{n,\bm{k}},\\-(1-{\rm{sin}}^2\theta)(1+\alpha E_0)+(1-{\rm{sin}}^2\theta+\alpha{\rm{sin}}^2\theta)\epsilon_{{\rm{F}},n,\bm{k}}&=&\alpha\epsilon_{n,\bm{k}},
\end{eqnarray}
\end{subequations}
leading to
\begin{equation}
\epsilon_{n,\bm{k}}=\sqrt{(2{\rm{sin}}^2\theta-1)\epsilon^2_{{\rm{F}},n,\bm{k}}+2E_0(1-{\rm{sin}}^2\theta)\epsilon_{{\rm{F}},n,\bm{k}}}.
\label{EoCAF}
\end{equation}

\begin{figure}[!htb]
\centering
  \includegraphics[clip=true,width=0.7\textwidth]{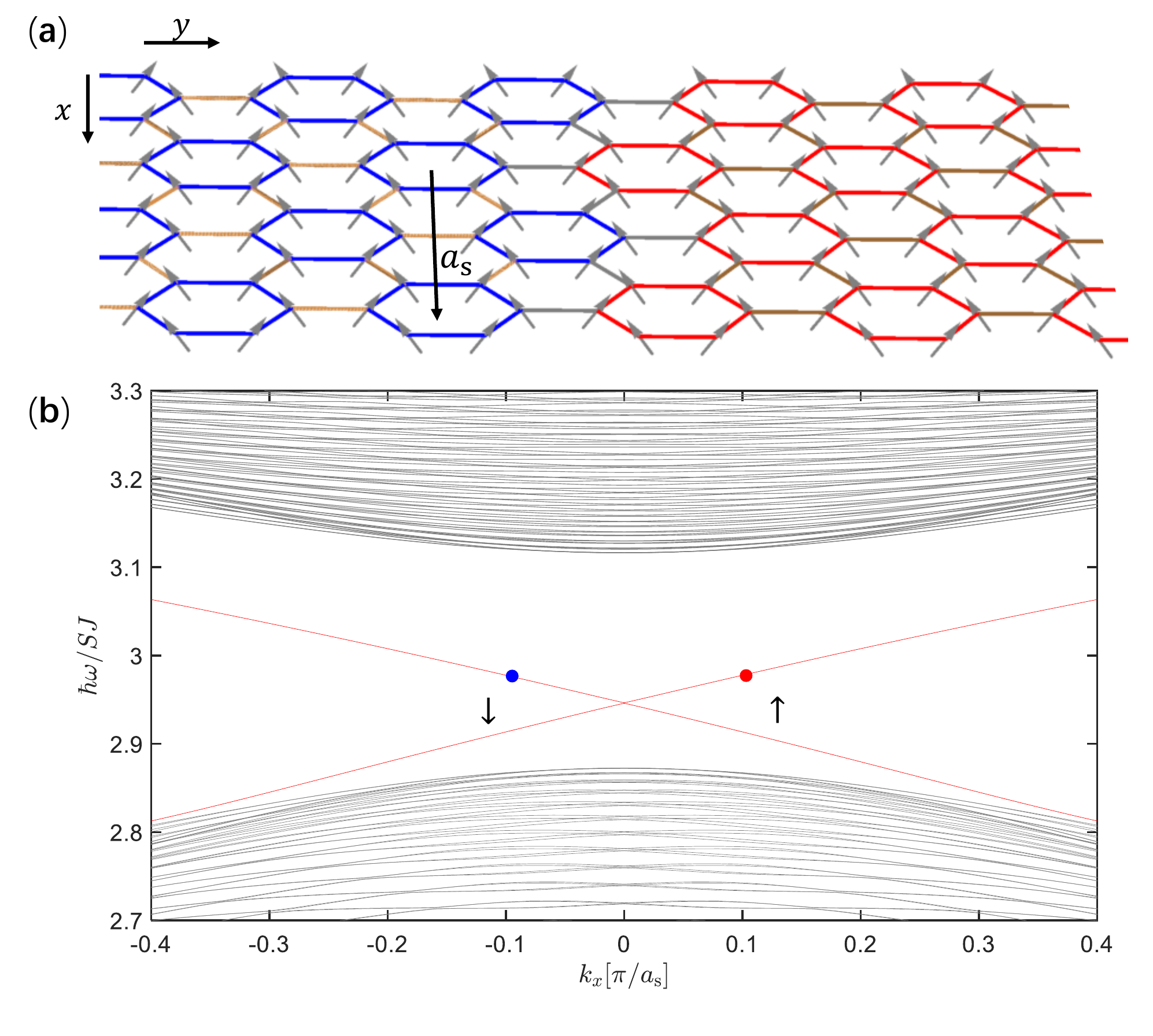}
  \caption{(a) Canted antiferromagnetic heterostructure containing a trivial domain and a topological domain, which is uniform and infinitely long in the $x$ direction. In the $y$ direction, 60 unit cells are contained in both trivial and topological domains periodically, where the exchange couplings are the same as in Figs. \ref{Fig:6}(b) and \ref{Fig:6}(d) respectively, which are denoted in the same way as in Fig. \ref{Fig:3}(a), and the couplings between the trivial and topological domains (denoted by gray lines) are given by geometric mean of $J_1$. (b) Frequency band structure of magnon modes for the heterostructure, where topological interface dispersions (red lines) appear in the bulk band gap.}
  \label{Fig:7}
\end{figure}

\begin{figure}[!htb]
\centering
  \includegraphics[clip=true,width=0.6\textwidth]{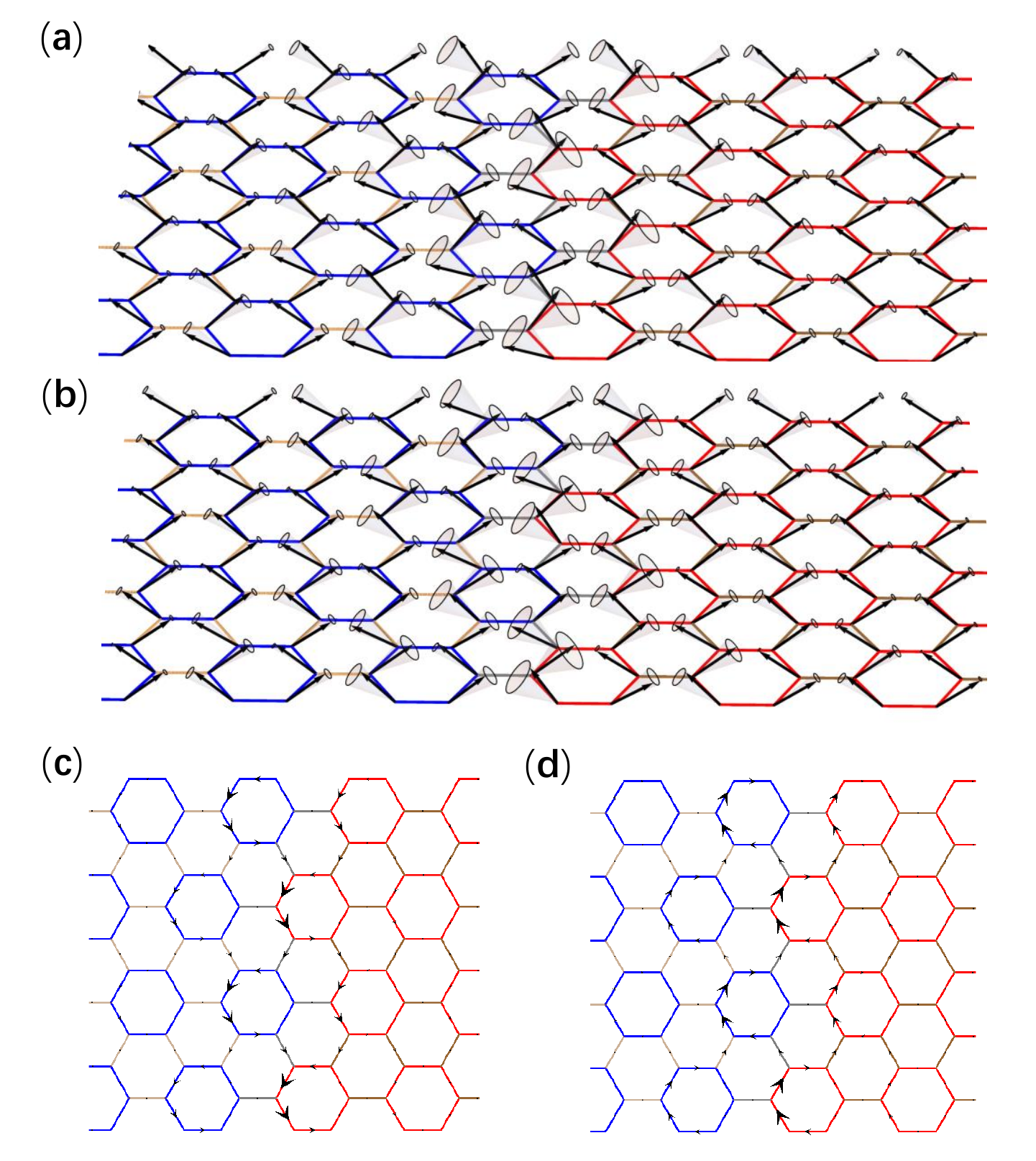}
  \caption{(a) and (b) Distribution of magnon modes at the two opposite momenta denoted by $\uparrow$ and $\downarrow$ in Fig. \ref{Fig:7}(b) respectively. (c) and (d) Same as (a) and (b) except for current distributions of magnon currents between sites presented by black arrows. The magnon mode denoted by $\uparrow$/$\downarrow$ carry up-/down-pseudospin, showing counterclockwise/clockwise magnon currents in unit cells.}
  \label{Fig:8}
\end{figure}

For ${\rm{sin}}^2\theta>1/3$, $\epsilon_{n,\bm{k}}$ is monotonic with $\epsilon_{{\rm{F}},n,\bm{k}}$. The $p$-$d$ band inversion is achieved in canted antiferromagnets by tuning coupling texture the same as in the ferromagnetic case. For ${\rm{sin}}^2\theta=0.4$, the full frequency band structures of canted antiferromagnets with the coupling textures are derived as shown in Figs. \ref{Fig:6}(b)-(d), and $p$ and $d$ magnon modes in canted antiferromagnets are shown in Figs. \ref{Fig:6}(e). For ${\rm{sin}}^2\theta<1/3$, the present scheme does not apply straightforwardly.

In order to see the topological interface magnon modes, we consider a heterostructure shown in Fig.~\ref{Fig:7}(a) similar to Fig. \ref{Fig:3}(a) for the ferromagnet. Topological interface dispersions appear inside the band gap of the frequency band structure as shown in Fig. \ref{Fig:7}(b). Magnon modes at the two momenta marked by dots in Fig. \ref{Fig:7}(b) are depicted in Figs. \ref{Fig:8}(a) and \ref{Fig:8}(b) respectively. We can calculate the magnon current for particle and hole parts of canted antiferromagnet using the same way of ferromagnetic case. The total magnon current is the magnon current of the particle part minus that of the hole part. Similar to topological interface magnon modes of the ferromagnetic case, in Figs. \ref{Fig:8}(c) and \ref{Fig:8}(d) magnon current circulates counterclockwise/clockwise in unit cells dominated by the up-/down-pseudospin, which also governs the direction of net magnon currents, demonstrating the pseudospin-momentum locking phenomenon in these topological interface magnon modes. Because the external magnetic field induces a ferromagnetic component $S\rm{sin}\theta$ on each site, the magnitude of magnon current is proportional to $\sin^2\theta$.

\section*{Discussion}

We propose a method to achieve topological magnon modes in magnetic systems on honeycomb lattice, including both ferromagnet and antiferromagnet. The frequency band structures are gapless for uniform nearest-neighbor exchange couplings. In ferromagnets, a topological frequency gap opens when exchange couplings inside the hexagonal unit cells are smaller than exchange couplings between unit cells, associated with a $p$-$d$ band inversion at the $\Gamma$ point. In antiferromagnets, the degeneracy in the frequency band structure due to the combination of time-reversal symmetry and inversion symmetry has to be lifted by applying an external magnetic field. The resulting canted antiferromagnets become topological upon tuning exchange couplings inside and between hexagonal unit cells same as in ferromagnets. Magnon currents of topological magnon modes propagate along the interface between a trivial domain and a topological domain in opposite directions governed by pseudospins, the circulation direction of magnon current in unit cells, manifesting the pseudospin-momentum locking phenomenon. In the future, candidate materials should be found to realize the scheme proposed in the present work. Besides specific materials, one may also observe topological magnon modes in artificial systems which can be achieved by depositing magnetic atoms on a metallic substrate using the STM technique or trapping magnetic atoms in an optical lattice using laser beams.

\section*{Methods}

The effective Hamiltonians of magnon modes in ferromagnet and antiferromagnet on honeycomb lattice are obtained by the Holstein-Primakoff approach. Besides, a local rotating frame and the Bogoliubov transformation are adopted for canted antiferromagnet. Frequency band structures and magnon modes are obtained by direct diagonalizations of Hamiltonian.

\normalem

\section*{Acknowledgements}

This work is supported by CREST, JST (Core Research for Evolutionary Science and Technology, Japan Science and Technology Agency), Grant No. JPMJCR18T4.

\section*{Author contributions statement}

X.H. supervised the project. H.H. conducted the numerical calculation. T.K. joined the discussions and writting of manuscript. All authors contributed to the discussions and the preparation of the manuscript.

\section*{Additional information}

\textbf{Competing interests}: The authors declare that they have no competing interests. 

The corresponding author is responsible for submitting a \href{http://www.nature.com/srep/policies/index.html#competing}{competing interests statement} on behalf of all authors of the paper. This statement must be included in the submitted article file.

\end{document}